\documentclass[aps,prc,twocolumn,amssymb,amsmath,superscriptaddress,floatfix,showpacs,nofootinbib]{revtex4}
\usepackage[dvips]{graphics,graphicx}
\usepackage{multirow}
\usepackage{longtable}
\begin{document}
\title{In-medium modification of $P$-wave charmonia from QCD
sum rules}
\author{Young-Ho Song}
\email{youngho.singer@gmail.com}
\author{Su Houng Lee}
\email{suhoung@phya.yonsei.ac.kr}
\author{Kenji Morita}
\email{morita@phya.yonsei.ac.kr}
\affiliation{Institute of Physics and Applied Physics, Yonsei
University, Seoul 120-749, Korea}
\date{\today}
\begin{abstract}
We investigate the changes of the masses and widths of $\chi_{c0}$ and
 $\chi_{c1}$ in hot gluonic matter near $T_c$ and in nuclear medium using QCD sum
rules.  As in the previous works for the $J/\psi$ and
 $\eta_c$, in-medium effects are incorporated through the changes of
 gluon condensates. Twist-2 terms for the $^3 P_0 (\chi_{c0})$  and
$^3 P_1 (\chi_{c1})$ are also included for the first time.
 The results show that larger mass shifts and width broadenings take place
as compared to the $S$-wave states.   As the critical change take place near $T_c$,  related measurements can reveal critical phenomenon in QCD.
\end{abstract}

\maketitle

\section{Introduction}

Spectral property of charmonia in medium can provide experimental
information about the confinement nature of QCD. At high temperature,
hadronic matter undergoes a phase transition into the deconfined phase.
Recent experimental data measured at Relativistic Heavy Ion Collider
(RHIC) reveals that the deconfined matter is strongly interacting \cite{BRAHMS_Whitepaper,PHOBOS_Whitepaper,PHENIX_Whitepaper,STAR_Whitepaper}.
It was previously expected that the production of $J/\psi$  will be suppressed by the color screening in the deconfined matter formed in the initial stages of heavy ion collisions  \cite{hashimoto86,Matsui_PLB178}.  On the other hand, recent experimental observation shows non-trivial suppression pattern that are not easily understood by a simple suppression mechanism.  An important fact in $J/\psi$ production is that
they are not only produced directly but also through decay from higher mass
states such as $\psi'$ and $\chi_{c}$. Since such excited states are likely to dissolve at just above the critical temperature $T_c$ whereas the  ground state does at 1.6$T_c$, as have been shown by lattice calculations  \cite{Asakawa_PRL92,Datta_PRD69,aarts07}, sequential suppression
scenario \cite{Karsch06} has been proposed, in which the initial suppression comes from the disappearance of the feed down from  $\psi'$ and $\chi_c$  followed by the suppression of  directly produced $J/\psi$.
In previous works, two of us (K.M and S.H.L) have studied  the properties of $J/\psi$ and
$\eta_c$ using QCD sum rules,  and have showed that their masses and widths
change following the critical behavior of gluon condensates, which
are extracted from lattice QCD \cite{morita_jpsiprl,morita_jpsifull}.
In this paper, we apply the same formalism to the $\chi_{c0}$  and $\chi_{c1}$ states to investigate how the properties of these states are affected by the QCD phase transition,  and explore the consequences in the experimental measurements. We also study the nuclear matter case which may
serve as a testing ground for the phase transition through precursor phenomena.

The paper organized as follows; in the next section, we
briefly summarize our formalism and present new Wilson coefficients.
In section \ref{sec:results}, we present quantitative results of in-medium
change of masses and widths of $\chi_{c0}$ and $\chi_{c1}$. Section
\ref{sec:discussion} is devoted to discussion and summary.

\section{Formalism}

QCD sum rules are based on the current correlation function and
the dispersion relation \cite{Shifman_NPB147}. In this paper, we consider the scalar current
$j^S=\bar{c}c$ and axial vector current
$j^A_\mu =(q_\mu q_\nu /q^2 -g_{\mu\nu})\bar{c}\gamma^\nu \gamma_5 c$
for $\chi_{c0}$ and $\chi_{c1}$, respectively. For finite temperature
we start with the  correlation function
\begin{equation}
 \Pi^J(q) = i\int d^4 x \, e^{iq\cdot x}
  \langle  T j(x) j(0)  \rangle_{T},
\end{equation}
with $\langle \cdots \rangle_{T}$ being the Gibbs average
$\langle \mathcal{O}(x) \rangle_T = \text{Tr}(e^{-\beta
H}\mathcal{O})/\text{Tr}(e^{-\beta H})$.   For the scalar, there is only
one independent polarization function, but for the axial vector
correlation, there are two independent polarization with respect to the
medium.  This is so because when the axial vector meson is moving with a
finite three momentum $\boldsymbol{q}$ with respect to the medium at
rest, the response will be different depending on whether the
polarization is parallel or perpendicular to the three momentum.
Hereafter, we assume that both the
medium and the $\bar{c}c$ pair are at rest so that
$q=(\omega,\boldsymbol{0})$.
Then, for both the scalar and axial vector meson, there will be only one
independent mode.  For the axial vector correlation function, one can
extract the  scalar correlation function through
$\tilde{\Pi}^{S}(q) = \Pi^S(q)/q^2$ and
$\tilde{\Pi}^A(q) = -\Pi^{A\mu}_\mu/(3q^2)$. Taking $q^2 < 0$, the
dispersion relation relates the correlation function with the spectral
density $\rho(s)$ \cite{Hatsuda93},
\begin{equation}
 \tilde{\Pi}^J(\omega^2)= \int_{0^-}^{\infty} du^2
  \frac{\rho(u)}{u^2-\omega^2}.
\end{equation}
In the QCD sum rules,
we calculate the correlation function by means of the operator
production expansion (OPE) with condensates while we model the
\textit{hadronic} spectral density with a pole
and continuum. In the present calculation, we neglect the continuum part
in the hadronic spectral density since we can suppress the contribution
by means of the moment sum rules.
Recently the role of the zero mode in the spectral density has been
emphasized in several literature \cite{umeda07,alberico08} since
it gives a constant contribution to the imaginary time correlator from
which the spectral density is reconstructed via maximum entropy method in
lattice QCD.  This comes from Landau damping effect, \textit{i.e.},
scattering of the current with on-shell quarks in medium and
has been known as the scattering term \cite{Bochekarev86} in the
framework of QCD sum rules.  This term is also related to the transport
coefficients as they are the terms responsible for the long range order
surviving at zero momentum and energy of the current. The scattering
term exists whenever the thermal excitation, thermal charm quark in our
present case, couples directly to the current, and contributes to the
spectral density at zero energy in the kinematical limit of
$\boldsymbol{q} \rightarrow 0$ with a strength proportional to the
thermal charm quark density at finite temperature
\cite{Furnstahl_PRD42}.  It can be shown that the same  scattering term
can be obtained in the OPE side by resuming the thermal charm quark
contribution to the zeroth order charm quark operators, that did not
have direct charm quark contribution at zero temperature as they were
converted to gluonic operators in the heavy quark
expansion~\cite{Generalis84}.  Therefore, while the scattering term
contribution is non-negligible in the spectral density, similar term in
the OPE will cancel its contribution and thus will not be considered
further in this work.
%Therefore, while the scattering term contribution is non-negligible and contribute to the spectral density at zero energy, its effect is irrelevant to the $J/\psi$ pole.  Similarly, it is also evident, that the changes of thermal gluon
%operators obtained in the quenched approximation, do not have much relation to scattering contributions in the phenomenological side and
%should dominantly influence the pole contribution.
In a sense, it is the advantage of the QCD sum rule method to artificially turn off the effects of thermal charm quarks altogether and concentrate on the thermal gluonic effects on the pole.
 For confined phase, there is no on-shell quarks and we do not have to
consider the scattering term from the beginning since we have only glueballs other than the heavy quarkonia in this phase.  Therefore, we will neglect the scattering terms in both the confined and deconfined phase.
%should be generally
%taken into account,
%we neglect it by focusing on the pure gluonic medium and nuclear matter
%in which no charm quarks can scatter with the current
%\cite{morita_jpsiprl,morita_jpsifull}.

For heavy quarkonia, the moment sum rule gives a systematic procedure. The
moment is defined as
\begin{equation}
 M_n(Q^2) = \left.\frac{1}{n!}\left( \frac{d}{dq^2} \right)^n
	     \tilde{\Pi}^J(q^2) \right|_{q^2 = -Q^2}.\label{eq:moment}
\end{equation}
Noting that the expansion parameter in the OPE for heavy quarkonia is
$\Lambda_{\text{QCD}}^2/(4m_c^2+Q^2)$ and focusing on temperature as large as
$\Lambda_{\text{QCD}}$ so that the expansion parameter is not much modified from the vacuum value, we can truncate the OPE at the leading order
perturbative correction and the dimension four gluon condensates, with the dimension four twist-2 gluon condensates being a new addition as compared to the vacuum case.  In this approach, all the temperature effects in the OPE are encoded in the temperature dependence of operators.
Therefore the moment \eqref{eq:moment} for the OPE side becomes, for $\xi=Q^2/4m_c^2$,
\begin{equation}
 M_n(\xi)=A_n(\xi)[1+a_n(\xi)\alpha_s(\xi)+b_n(\xi)\phi_b+c_n(\xi)\phi_c],\label{eq:opemoment}
\end{equation}
where $A_n$, $\alpha_n$, $b_n$ and $c_n$ are the Wilson coefficients for
the bare loop, perturbative radiative correction, scalar gluon
condensate and twist-2 gluon condensate, respectively. In
Eq.~\eqref{eq:opemoment}, $\phi_b$ and $\phi_c$ have temperature
dependencies through the gluon condensates. They have been extracted
from lattice calculation of pure SU(3) theory and shown in
Ref.~\cite{morita_jpsifull}. Here we would like to stress that they are
not order parameters in a strict sense but their behavior more or less reflect
the phase transition, \textit{i.e.}, sudden change across the critical
temperature $T_c$.

As for the Wilson coefficients, $A_n$, $a_n$ and $b_n$ are listed in
Ref.~\cite{Reinders_NPB186}. $c_n$ for the scalar and axial currents are
calculated for the first time in this work by making use of the
background field technique \cite{novikov84_fortphys}. Defining the
scalar functions
$\tilde{\Pi}^S (q) = \Pi^S(q)/q^2$ and
$\tilde{\Pi}^A(q)=-\frac{1}{3q^2}\Pi^{A\mu}_\mu$
for the scalar channel and the axial vector channel correlation
function, respectively, we find that the correction term for the twist-2
gluon operators in OPE becomes
\begin{widetext}
\begin{align}
 \Delta \tilde{\Pi}^S(q)&= \left\langle
 \frac{\alpha_s}{\pi}G^a_{\alpha\mu}G^{a\beta\mu} \right\rangle
 \frac{q^\alpha q_\beta}{Q^2}
 \frac{1}{Q^4}\left[ \frac{3}{2}J_2(y)-\left( 1-\frac{1}{3}y
 \right)J_1(y) -\frac{1}{2}
 \right]\\
 \Delta \tilde{\Pi}^A(q)&= \left\langle
 \frac{\alpha_s}{\pi}G^a_{\alpha\mu}G^{a\beta\mu} \right\rangle
 \frac{q^\alpha q_\beta}{Q^2}
 \frac{1}{6Q^4} \left[6J_2(y)+\frac{4}{3}y J_1(y)-6\right],
\end{align}
\end{widetext}
where $y=Q^2/m^2$ and $J_n(y)=\int_{0}^{1}dx [1+x(1-x)y]^{-n}$.
Therefore $c_n$ is given by
\begin{align}
 c_n^S(\xi)&=
 b_n^S(\xi)+\frac{4n(n+1)}{3(1+\xi)}\frac{F(n+1,\frac{1}{2},n+\frac{5}{2};\rho)}{F(n,\frac{3}{2},n+\frac{5}{2};\rho)},\\
 c_n^A(\xi)&=  b_n^A(\xi)+\frac{4n(n+1)}{3(1+\xi)}\frac{F(n+1,\frac{1}{2},n+\frac{5}{2};\rho)}{F(n,\frac{3}{2},n+\frac{5}{2};\rho)},
\end{align}
where $\rho=\frac{\xi}{1+\xi}$ and $F(a,b,c;z)$ denotes the hypergeometric
function $_2F_1(a,b,c;z)$. One can see that $c_n$ can be expressed
as a sum of $b_n$ and an additional term, as in the case of the vector and
pseudoscalar case \cite{Klingl_PRL82}, and that the additional terms of
the scalar and axial vector channel are identical.

Masses and widths are extracted in the same manner as in the previous works
for $J/\psi$ and $\eta_c$ \cite{morita_jpsiprl,morita_jpsifull}. Taking
ratio of the moments and equating the OPE side with the phenomenological
side, we find
\begin{equation}
 \left.\frac{M_{n-1}}{M_n}\right|_{\text{OPE}} = \left.\frac{M_{n-1}}{M_n}\right|_{\text{phen.}}.\label{eq:ratio}
\end{equation}
In the phenomenological side, we simply express the pole term with the
relativistic Breit-Wigner form and neglect the continuum contribution.
Namely, we put
\begin{equation}
 \rho(s) = \frac{1}{\pi}\frac{f_0 \Gamma \sqrt{s}}{(s-m^2)^2+s\Gamma^2}.
\end{equation}
This simplification is justified by taking the moment since it
suppresses the high energy contribution in the phenomenological side.
While the continuum slightly modifies the mass value, its contribution does not affect the in-medium change from the vacuum value, which is the interest of this paper.
Also, the thermal factor $\tanh(\sqrt{s}/2T)$ is put to unity since we are
considering temperatures much lower than the mass of charmonium.

\section{Results}
\label{sec:results}

\begin{figure}[ht]
 \includegraphics[width=3.375in]{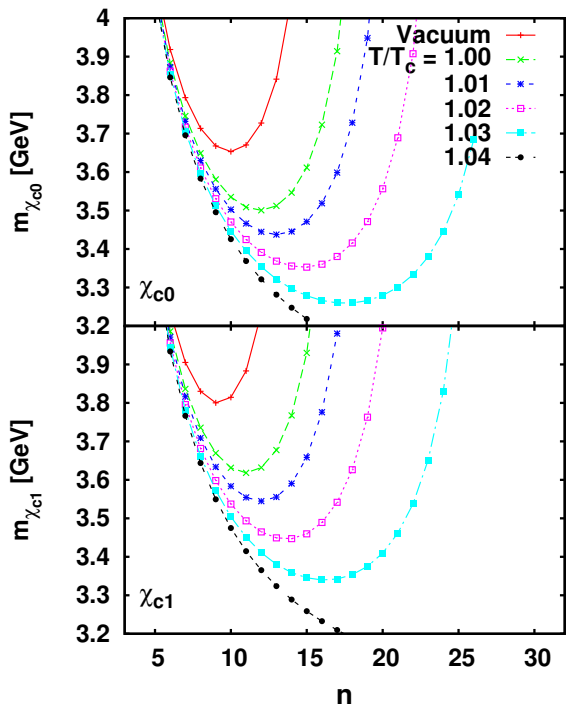}
 \caption{(Color online). Masses in the $\Gamma \rightarrow 0$ limit as
 a function of $n$ and $T/T_c$. The upper panel
 is for $\chi_{c0}$ and the lower for $\chi_{c1}$.}
 \label{fig:momentratio}
\end{figure}
Following Ref.~\cite{Reinders_NPB186}, we work with $\xi=2.5$ to validate
the OPE since the Wilson coefficients are larger than those of the $S$-wave.
The parameters are the same as in Ref.~\cite{morita_jpsiprl}, with
$\alpha_s(\xi=1)=0.21$ and $m_c(\xi=1)=1.24$ GeV.
Since we have neglected the continuum, vacuum masses are a little larger
than the experimental data. However, it does not affect the results of
medium-induced changes discussed in what follows, as we will discuss later.

\begin{figure}[ht]
 \includegraphics[width=3.375in]{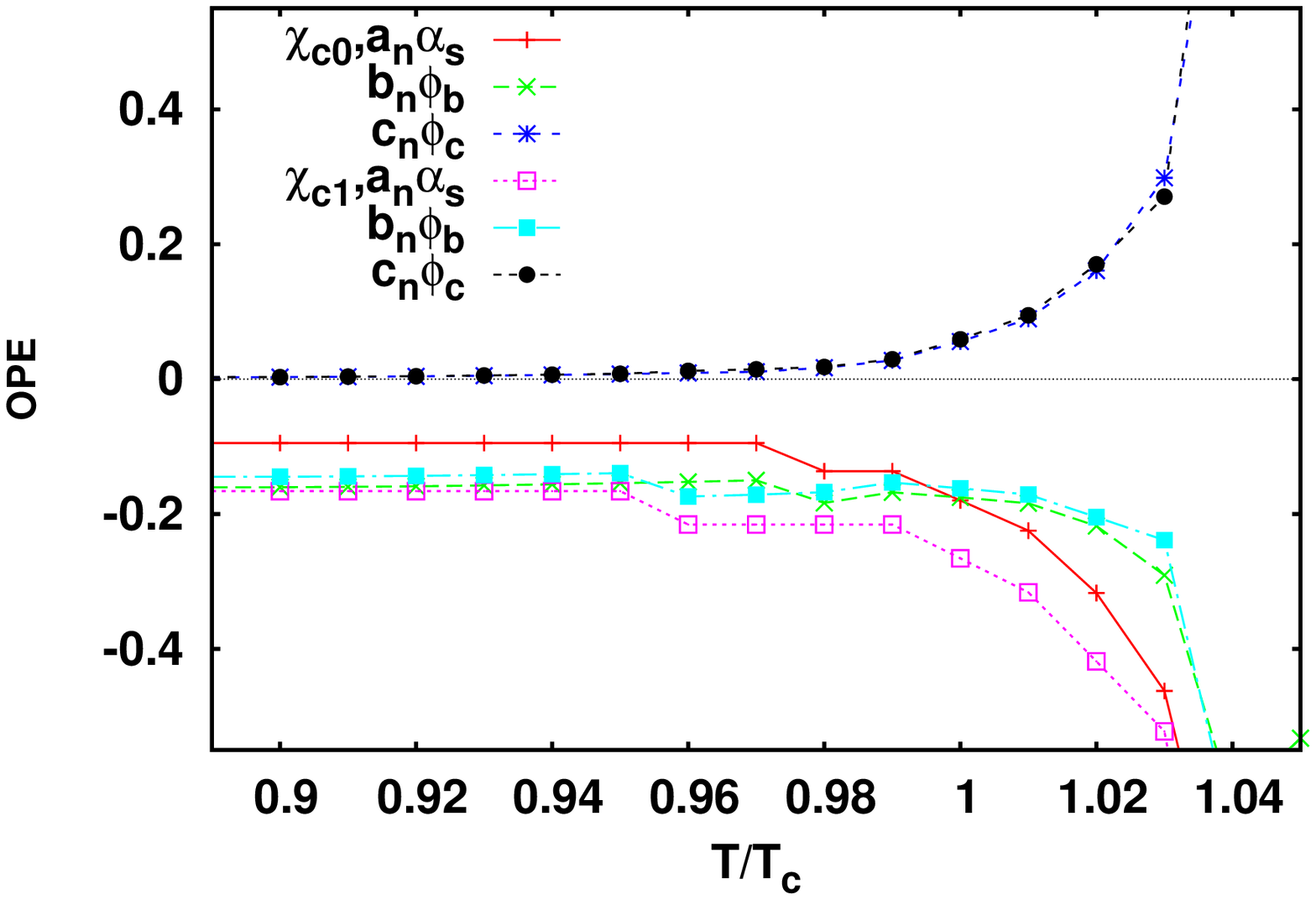}
 \caption{(Color online). OPE terms as a function of $T/T_c$. At each temperatures, the
 terms are calculated at $n$ which gives the minimum of the moment ratio
 as shown in Fig.~\ref{fig:momentratio}.}
 \label{fig:opevstemperature}

 \includegraphics[width=3.375in]{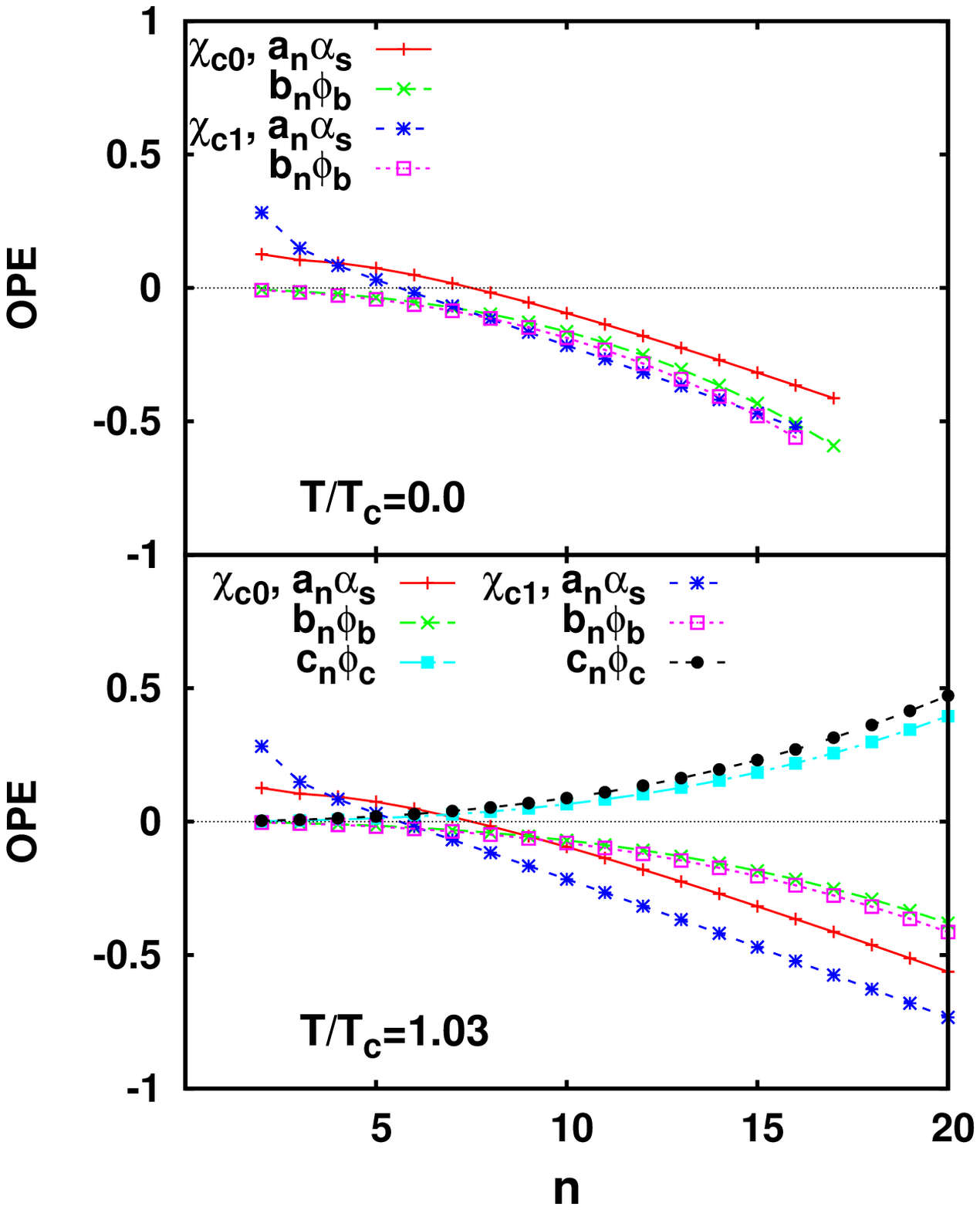}
 \caption{(Color online). OPE terms as a function of $n$ for vacuum and $T=1.03T_c$.}
 \label{fig:opevsn}
\end{figure}

Figure \ref{fig:momentratio} shows the masses in the
$\Gamma \rightarrow 0$ limit calculated from
$m^2 =M_{n-1}/M_n|_{\text{OPE}}-4m_c^2 \xi$
at various temperatures. At large $n$, the OPE breaks down, while at
small $n$, the continuum contribution becomes important.  Therefore, the
region where both approximations are good enough and the sum rule is reliable
shows up in the plot as a minimum, the value at which corresponds to the
physical mass \cite{Reinders_NPB186,Narison_sumruletextbook}. As seen
in the $S$-wave cases, the masses decrease as temperature increases. The
sum rule can be applied as long as the moment ratio has a minimum. We
can see that the OPE breaks down at $T > 1.03T_c$ for both $\chi_{c0}$ and
$\chi_{c1}$. To see the behavior clearly, we also display each term
in Eq.~\eqref{eq:opemoment} in Figs.~\ref{fig:opevstemperature} and \ref{fig:opevsn}.
One can see that the OPE breaks down because of large expansion
coefficients which should be smaller than, typically, 0.3. At the
marginal temperature $T=1.03T_c$, the perturbative correction term is
close to 0.5; this is so because the minimum of the moment ratio is shifted to
the large $n$ direction where the correction terms increase as can be seen in Fig.~\ref{fig:opevsn}.  Therefore, the result at this temperature is less reliable than that at lower temperatures.
 From Fig.~\ref{fig:opevsn}, one also notes the opposite
sign between $b_n\phi_b$ (scalar gluon condensate) and $c_n \phi_c$
(twist-2). This is a general feature of the condensates contribution
coming from the fact that the scalar condensate remains positive at temperatures not so
higher than $T_c$ while the twist-2 operator is always negative.
Their magnitudes almost coincide at $T=1.03T_c$ and the twist-2 term exceeds the
scalar condensate at $T=1.04T_c$.  The fact that the  individual OPE term is large, and that the perturbative $\alpha_s$ contribution becomes relatively larger than the scalar condensate, lead to the loss of stability of the OPE at the marginal temperature, whose convergence guarantees the existence of the bound state.
However, definite
conclusion on whether $\chi_c$ melts cannot be drawn
from the current consideration.

\begin{figure}[ht]
 \includegraphics[width=3.375in]{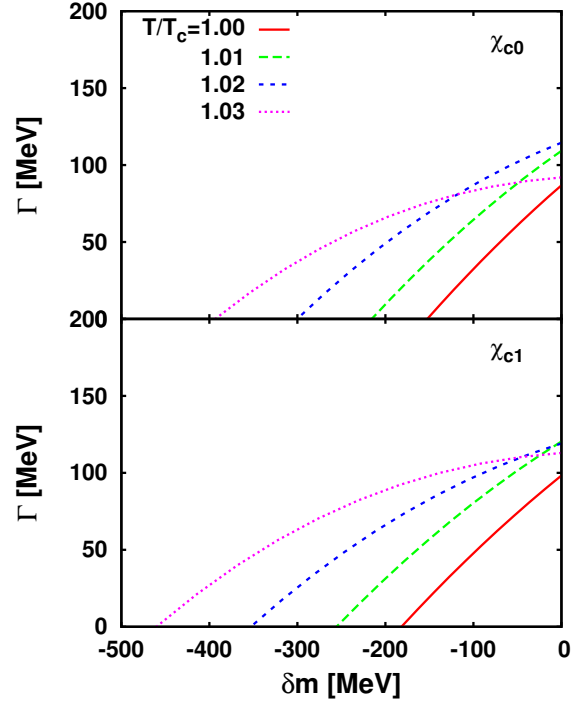}
 \caption{(Color online). Relation between the mass shift and width
 broadening at temperatures above $T_c$. The upper and lower panels show
 $\chi_{c0}$ and  $\chi_{c1}$ cases, respectively.}
 \label{fig:dm-g}
\end{figure}

Solving Eq.~\eqref{eq:ratio} with respect to $m$ and
$\Gamma$ for the minimum of OPE side moment ratio at each temperature,
we obtain the relation between mass shift and width broadening.
The results are displayed in Fig.~\ref{fig:dm-g}.
As in the $S$-wave cases, a linear relation is also seen in the
$P$-wave quarkonia. In the small mass shift and large broadening case at
$T=1.03T_c$, the relation shows some deviations from the linear
relation. This comes from the fact that the stability is achieved at
large $n$ which gives strong dependence of the moment ratio on $n$.
As mentioned before, convergence of OPE is not so good.
Strictly speaking, physical parameters extracted from the sum rules
should not strongly depend on the external parameters, $\xi$ and $n$.
In this case, however, $\Gamma$ at $\delta m =0$ depends on $n$ while
$\delta m$ does not. Therefore the results of the width might be less
reliable. In reality there should be mass shifts in this temperature
region caused by change of the chromo-electric field, \textit{i.e.},
second-order Stark effect \cite{lee_morita_stark,brambilla08}. Hence, a small mass
shift with large width broadening would not be a realistic combination.

\begin{figure}[ht]
 \includegraphics[width=3.375in]{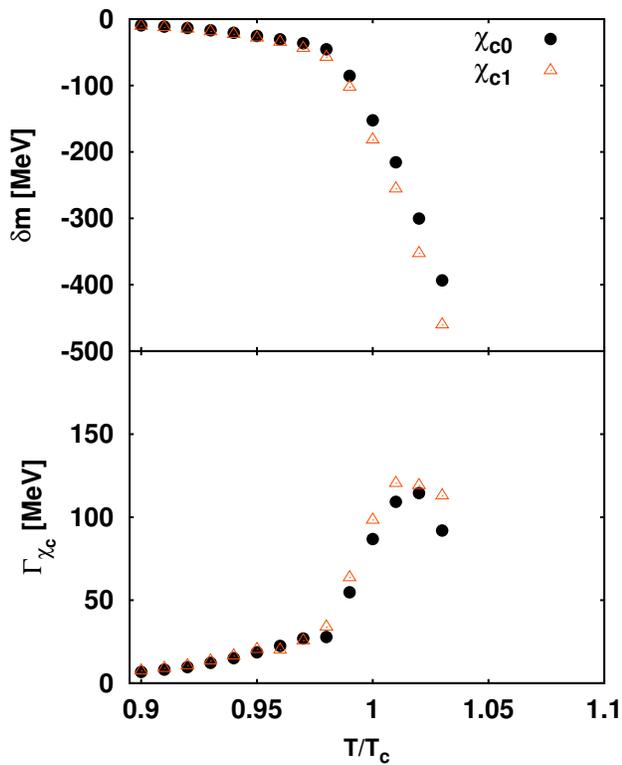}
 \caption{(Color online). Upper: mass shift as a function of
 $T/T_c$. Lower: Width broadening as a function of $T/T_c$. Circles and
 triangles denote the $\chi_{c0}$ and $\chi_{c1}$, respectively.}
 \label{fig:dmdg-T}
\end{figure}

To see the temperature dependence, we depict the temperature dependence of
the maximum mass shift and width broadening in Fig.~\ref{fig:dmdg-T}.
One can see that both $\chi_{c0}$ and $\chi_{c1}$ exhibit a clear
critical behavior with respect to the abrupt change of the gluon
condensates. This qualitative feature was also seen in the $S$-wave
\cite{morita_jpsifull} cases, but in the present case, the spectral
changes are roughly a factor of two larger . Especially the maximum mass
shifts reach
$\sim$150 MeV at $T=T_c$ and more than 400 MeV at $T=1.03T_c$.   Lattice calculation suggests~\cite{Datta_PRD69} that the $\chi_{c0}$ and $\chi_{c1}$ will dissolve before $T=1.1T_c$.  However, as is evident also in the lattice data, our result suggest that just before it dissolves, the spectral density will be greatly modified and the mass greatly reduced slightly above $T_c$, which are
more promising signals for direct observation in future
experiments than the correspondingly smaller mass shift expected for $J/\psi$.

\begin{figure}[ht]
 \includegraphics[width=3.375in]{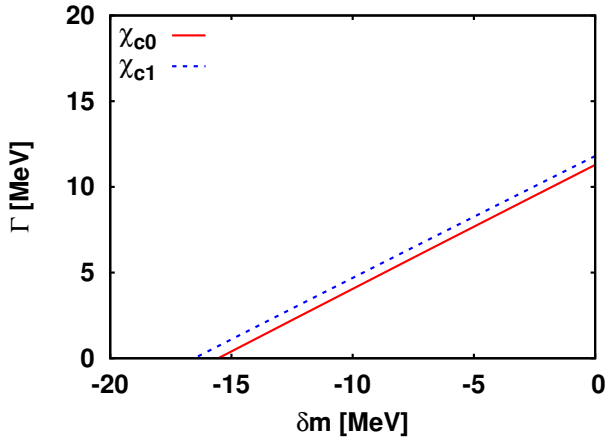}
 \caption{Relation between mass shift and width broadening in the
 nuclear matter. Solid and dashed lines stand for $\chi_{c0}$ and
 $\chi_{c1}$, respectively.}
 \label{fig:nm}
\end{figure}

We also calculate the mass shift and width broadening in nuclear matter.
This can be done by replacing the change of the gluon condensates with
those for the nuclear matter \cite{Klingl_PRL82,morita_jpsifull}. The
result of mass shift and width broadening is shown in Fig.~\ref{fig:nm}.
The maximum mass shift is around 17 MeV, which is more than two times
larger than that obtained previously for the $S$-wave charmonia.
The difference between $\chi_{c0}$ and $\chi_{c1}$ is around 1 MeV.

\section{Summary and discussions}
\label{sec:discussion}

In this paper, we extend the previous QCD sum rules analyses for the
$S$-wave charmonia to $P$-waves. The results are qualitatively similar to
those of $S$-waves but show larger spectral changes. Due to the
absence of convergence of the OPE, our approach is limited to
temperatures above but close to $T_c$. Note that our results do not
necessarily imply the absence of the bound states; spectral function
obtained in lattice calculations does not have a good resolution of the
width. Although potential models are extensively studied with potentials
extracted from lattice calculations \cite{Wong05,alberico05,alberico07,mocsy08},
it is not straightforward to implement the critical change into
potentials at fixed temperatures, and also ambiguities exist in the choice of the potential.
%At present, we do not have a consistent conclusion but some aspects
%seem to agree each other. Both our approach and potential models imply
%mass shifts and finite thermal width near $T_c$.

One may wonder if the continuum part in the hadronic
spectral function that we have neglect in this paper can affect the
result. As described before, contribution from
this part can be suppressed by making use of the moment sum rule prescription.
However, this suppression lead to too large $n$ of the moment at
$T \geq 1.03 T_c$,
which results in the non-monotonic behavior of the width in
Fig.~\ref{fig:dmdg-T}. We will be able to improve it by taking the
continuum into account. In this case, however, the continuum threshold
$s_0$ in the phenomenological side appears as a free parameter.
This can be fixed so that mass agrees with the experimental value
in the vacuum case, while we have no criteria in the case of medium.
This means the change of the OPE in medium might be attributed to the change of the threshold.
Nevertheless, present calculation has general consequences
because it corresponds to results at $s_0\rightarrow \infty$
limit. Therefore, Fig.~\ref{fig:momentratio} shows
the maximum mass at each temperatures, as it is well known in the charmonium sum rule \cite{Reinders_NPB186} that the extracted mass
increases with increasing $s_0$.
% One can realize that
%the masses are smaller than the experimental values.
Consequently,
if one adjusts $s_0$ so as to reproduce the experimental value at zero
temperature and then keep it
fixed as temperature changes, the analysis will result in similar amount of
spectral changes. If one allows $s_0$ to change with temperature, the mass shifts will be larger than those obtained at
$s_0 \rightarrow \infty$. Even in this case,
there are clear evidence of the change of the masses at $T > 1.02T_c$ at
which masses are smaller than experimental values.
Further discussion on the effect of the continuum will be given in a
future publication \cite{lee_morita_song_inprep}.  Future works also
include directly comparing the QCD sum rule result with the $T$-dependence
of the Euclidean correlation functions observed in
lattice QCD calculations.

In this paper, we utilized the gluon condensates extracted from lattice
gauge theory of the pure gluonic system. Since there are many differences
between the quenched and full QCD, one may wonder if the present
results are relevant for realistic situations. First, the effect of
dynamical quarks alters thermodynamic properties. Namely, transition
temperature reduces to 170-200 MeV \cite{cheng08} while it is around 260
MeV for pure SU(3) \cite{Boyd_NPB469}. On the other hand, what is important for the present analysis is the  the quantitative change in magnitude of the condensate near $T_c$.
%The gluon condensates, which are related to the thermodynamic quantities, changes a little by dynamical quarks.
Figure 21 of Ref.~\cite{morita_jpsifull} shows the behavior of the
scalar gluon condensates of full QCD case after the fermionic
contribution is subtracted out. One sees that the amount of decrease
from the vacuum value is almost the same as in the pure gauge case
while the change around $T_c$ becomes rather moderate. Considering the fact that the present lattice results have been performed on quite coarse lattice for which the hadron spectrum is still distorted and thus the thermodynamic quantities in the low temperature region are not properly calculated the changes of gluon condensate could be larger below $T_c$~\cite{Kharzeev07}.
This means that the spectral
changes set in at much lower temperature below $T_c$ due to the large
number of degree of freedom in the hadronic phase and the amount of
change at $T_c$ is almost the same.
Although not available at present, the twist-2 gluon part will also
change a little as it is proportional to the entropy density of the gluons,
which will not be effected greatly by the dynamical quarks. Hence, our
present result could change a little but the main features will remain if dynamical quarks are introduced.

The results contain rich physical consequences relevant
to recent and future experiments. First of all, a large fraction of $J/\psi$
comes from $\chi_c$. The in-medium modification of $\chi_c$ will affect
the final yields of $J/\psi$. If both  $J/\psi$ and $\chi_c$ can exist
slightly above $T_c$, $\chi_c$ with modified mass and width will decay into
$J/\psi$. This will give further enhancement of $J/\psi$ production.
Moreover, if the charmonia will be produced statistically \cite{andronic08} at the
same hadronization temperature, the larger mass shift of $\chi_c$ will lead to a larger enhancement of the $\chi_c$ compared to $J/\psi$ and then will result in the change of the production ratio $N_{J/\psi}/N_{\chi_c}$ in nucleus-nucleus collisions.
Alternatively, since the width of $\chi_c$ can be large enough to decay
inside medium, it might provide a chance to directly measure the large
mass shift. This is experimentally a challenging subject,
which requires the reconstruction of the $J/\psi+\gamma$ invariant mass distribution
with extremely fine resolution, but if successful
will provide a direct evidence of the critical phenomenon in QCD.
Our result for the nuclear matter can be a testing ground of this idea.
The maximally expected mass shift around 15 MeV may not be large, but
enough to be observable in the anti-proton project at FAIR~\cite{Lee04}.
This will be another challenge for future experiments.

To conclude, we have studied the in-medium spectral change of
$\chi_{c0}$ and $\chi_{c1}$ with the QCD sum rule approach developed in
Ref.~\cite{morita_jpsiprl}. Our results show that the change in these
states are more prominent than previously studied $S$-wave states, and this result will provide another means to probe critical phenomena in QCD.

\section*{Acknowledgment}
This work was supported by BK21 program of the Korean Ministry of
Education. S.H.L. was supported by the Korean Research Foundation
KRF-2006-C00011 and by the Yonsei University research fund.

%\bibliography{charm,chiral,eos,experiment,hydro,jpsi_sup,lattice,morita,nuclearmatter,QGPreview,sumrule,textbook,thermalmodel}

\end{document}